\documentclass[conference]{IEEEtran}

\IEEEoverridecommandlockouts

\usepackage{cite}
\usepackage{graphicx}
\usepackage{adjustbox}
\usepackage{eurosym}
\graphicspath{ {images/} }
\usepackage{amssymb}
\usepackage{amsthm}
\usepackage{amsmath}
\DeclareMathOperator*{\argmax}{arg\,max}
\usepackage{verbatim}%for begin comment

\usepackage{algorithm}
\usepackage[noend]{algpseudocode}

\algdef{SE}[SUBALG]{Indent}{EndIndent}{}{\algorithmicend\ }%
\algtext*{Indent}
\algtext*{EndIndent}

\usepackage{url}

\begin{document}

\title{Gibbs sampling for game-theoretic modeling of private network upgrades with distributed generation}
%\title{Game-theoretic Modeling of Curtailment Rules and their Effect on Transmission Line Investments}
%\title{Private network upgrade with distributed generation}

\author{\IEEEauthorblockN{Merlinda Andoni,
Valentin Robu, David Flynn and Wolf-Gerrit Fr{\"u}h}
\IEEEauthorblockA{Heriot-Watt University, Edinburgh, UK\\
Email: $\{$m.andoni,v.robu,d.flynn,w.g.fruh$\}$@hw.ac.uk}}

%\IEEEpubid{978-1-5386-4505-5/18/\$31.00 ~\copyright˜2018 IEEE}

\maketitle

\begin{abstract}
Renewable energy is increasingly being curtailed, due to oversupply or network constraints. Curtailment can be partially avoided by smart grid management, but the long term solution is network reinforcement. Network upgrades, however, can be costly, so recent interest has focused on incentivising private investors to participate in network investments. In this paper, we study settings where a private renewable investor constructs a power line, but also provides access to other generators that pay a transmission fee. The decisions on optimal (and interdependent) renewable capacities built by investors, affect the resulting curtailment and profitability of projects, and can be formulated as a Stackelberg game. Optimal capacities rely jointly on stochastic variables, such as the renewable resource at project location. In this paper, we show how Markov chain Monte Carlo (MCMC) and Gibbs sampling techniques, can be used to generate observations from historic resource data and simulate multiple future scenarios. Finally, we validate and apply our game-theoretic formulation of the investment decision, to a real network upgrade problem in the UK.
\end{abstract}

% \fi
\IEEEpeerreviewmaketitle

\section{Introduction}
%\IEEEpubidadjcol
Renewable energy sources (RES) play a key role in the climate change mitigation agenda. RES are variable, depend on weather patterns and are difficult to predict, hence raise technical challenges regarding network management. Moreover, grid infrastructure is often inadequate to support RES development, especially in the area of distribution networks. For instance, often RES projects are clustered in remote areas of the grid, where planning approval may be favourable and renewable resources abundant. Typically, in the UK, such areas are windy islands with constrained connections to the main grid.%, facing network constraints, such as voltage, frequency or fault level violations in the absence of a network upgrade. Examples include the Shetland and Orkney archipelagos and the Kintyre peninsula, used as a case study to validate the model developed in this work.
~In these areas, technical limitations and imbalance of renewable supply to demand, often result in RES curtailment, i.e. the energy that could have been generated is wasted as the system cannot absorb it or transfer it where required. %Typically areas increasingly affected by curtailment are areas of excess RES generation, areas where at times local demand is very low when compared to RES generation or simply areas with limited or saturated connection to larger areas of the grid.
~Curtailment can affect implementation of future RES projects, result in lost revenues and instigate socio-economic challenges, especially when local community-owned projects are involved.

Typically, RES generators are granted firm connections to the grid and receive compensation for incurred curtailment, the cost of which is eventually borne by all system users.%In many countries generators have been compensated when curtailment occurs however, part of costs are bared by all consumers, (approximately 18\% of the average electricity bill of a typical UK household \cite{ward2012demand} represents system charges for example) therefore 
~However in several occasions, RES generators are offered interruptible, non-firm connections, as an alternative to expensive or time consuming reinforcements. The so-called flexible commercial arrangements are increasingly offered by network operators, as an alternative, and are creating a shift in network access rules. A detailed review on flexible commercial arrangements can be found in~\cite{estopier2013flexible, kane2014review}. As shown in~\cite{andoni2016game}, arrangements that are fair and equally share curtailment and access among generators, can maximise the generation capacity built at a certain location and minimise discouragement for future investors.

Solutions to reduce curtailment include demand side management, energy storage or smart grid techniques such as Dynamic Line Rating (DLR) or real-time monitoring of the thermal state of the lines~\cite{michiorri2011dynamic} and Active Network Management (ANM), i.e., the automatic control of the power system by control devices and data that allow real time operation and optimal power flows~\cite{anaya2015options}.%From the DNO perspective, both techniques imply controlling DG power outputs, hence innovative commercial agreements between generators and the system operator are required.
~The long term solution, however, is network upgrade.%, such as reinforcing or building new transmission/distribution lines.
~Anaya \& Pollit (2015) compared smart interruptible connections to traditional grid reinforcements in~\cite{anaya2015options}.
~It is estimated that, by 2030 in the US alone, up to \$2 trillion will be required for network upgrades~\cite{bronski2015economics}. As grid expansion is expensive, it is desirable to provide incentives to private parties for taking part in such investments. ~Private network upgrades have been studied in several works~\cite{maurovich2015transmission, perrault2014efficient}.
~However, they raise the question for system operators of defining the framework within which these private lines are incentivised, built and accessed by competing generators.
\IEEEpubidadjcol

New private lines constructed, often follow a `single access' principle, i.e. lines for sole-use that suffice only to accommodate the RES capacity of each project. However, as shown in~\cite{andoni2016game}, it is possible for system operators to encourage RES generators to install larger capacity lines under a \emph{`common access'} principle, i.e. a private investor is licensed to build a line only if it grants access to smaller generators, which are subject to transmission charges. In these settings, curtailment and line access rules can play a significant role in the resulting grid expansion. Crucially, this leads to a leader-follower or a \textit{Stackelberg game between the line investor and local generators} respectively. Stackelberg games in network upgrades and RES settings have been presented in~\cite{van2012economics, asimakopoulou2013leader}.%, Locational Marginal Pricing \cite{shrestha2004congestion} or highlight the uncertainties of RES generation \cite{van2012economics}. Recent work in the renewable energy domain used Stackelberg game analysis to study energy trading among microgrids \cite{asimakopoulou2013leader, lee2015distributed}.
%In these settings, curtailment and line access rules can play a significant role in the resulting strategic expansion. Stackelberg equilibria are classified as solutions to sequential hierarchical problems where a dominant player (here the line investor) has the market power to impose their strategies to smaller players (local generators) and influence the price equilibrium. The equilibrium of the emerging game determines the optimal generation capacity installed and resulting profits for both players under varying cost parameters. A feasible range of the transmission fee is identified allowing both transmission and generation capacity investments be profitable.
~Our previous work was one of the first to introduce Stackelberg game formulations in settings that combine network upgrades, curtailment and line access rules, %introduced the formulation of the game-theoretic model,
~however the model presented in~\cite{andoni2016game} did not take into account stochastic generation and demand required for equilibrium estimation. Subsequent work~\cite{andoni2017game} improved by utilising real data,%to characterise the equilibrium of the game. 
~however, followed a one-shot, single scenario approach.% based on real data analysis.
~We build on this work by developing a principled framework, based on game-theoretic and state-of-the-art sampling techniques, i.e. Markov chain Monte Carlo (MCMC). Several authors used MCMC for modelling of wind speeds or wind power outputs~\cite{papaefthymiou2008mcmc, wu2012markov}. %\textbf{machine learning and big data techniques}. 
~Our framework allows modelling multiple renewable investment scenarios that reduce the uncertainty of future generation and demand.
~In more detail, the main contributions of the work are:
%\vspace{-1mm}
\begin{itemize}
\item We develop a new methodology that generates observations from renewable resource data. While historic data, such as wind speeds may be available, they might have considerable gaps and joint distributions cannot be expressed in simple, closed-form equations. For this reason we develop a MCMC methodology (Gibbs sampling) that can draw samples from available data and run multiple scenarios of potential futures.
%We utilise MCMC methods and more particular Gibbs sampling to run multiple scenarios of potential futures. We have formalised a new methodology that draws samples directly from wind data. The methodology can be used to draw samples from the unknown joint distribution of the wind speeds of generation locations. In addition, we can draw samples from the demand data to compute generation and resulting curtailment on on hourly basis.
%\vspace{-1mm}
\item We establish a methodology that can determine optimal generation capacity investments through use of real demand and wind speed data. This work is one of the first to combine Stackelberg equilibria to a large-scale realistic game with MCMC techniques. Our model designates players' actions, depending on RES output correlation and expected curtailment, and studies the cost parameters effects on the equilibrium of the game.
%\vspace{-1mm}
\item Previous work has been extended to account for varying demand data instead of an average approach followed in both~\cite{andoni2016game, andoni2017game}. The model is applied and validated in a real network upgrade problem in the UK.% Theoretical analysis is applied to a real network upgrade problem in the UK, based on real-data for wind speeds and demand that span over the course of 17 years. This case study analysis demonstrates how the methodology can be applied in practical settings.
%Finally, from historical data we can build distributions however curtailment depends on joint distribution. This joint distribution cannot be expressed analytically and is unknown. However our methodology can directly draw samples from this unknown distribution. 
\end{itemize}
%\vspace{-1mm}
Section~II of the paper presents the theoretical formulation of the game, Section~III introduces Gibbs sampling, Section~IV demonstrates the methodology based on a real-world case study, Section~V shows the results and Section~VI concludes.

\section{Stackelberg game model description}
\label{Gen_model}
%\textbf{In this section, we examine the combined effects that fair curtailment rules and `common access' line rules have on network upgrade and the renewable capacity installed. First, we describe the general setting of the game. Second, we present a model that captures stochasticity of generation and variation in demand.}

Consider a simplified two-node network: location~A is a net consumer (area of high demand)%(where demand exceeds supply, e.g.~a mainland location with industry or significant population density)
~and location B is a net producer (area of high wind resource).%renewable conditions, e.g.~a remote region rich in wind resource).%In practice, there would be some local demand and supply, considered here negligible, and installation of new RES capacity is not be feasible without a network upgrade.
~Moreover, consider two players: a line investor, who can be merchant-type or a utility company and is building i) the AB interconnection and ii) $\textstyle {P_{N_1}}$ renewable capacity at B, and a local player representing all local RES generators located at B, and who builds renewable capacity equal to $\textstyle {P_{N_2}}$. This second player can be thought of as investors from the local community, who do not have the technical/financial capacity to build a line, but may have access to cheaper land, find it easier to get social approval to build turbines etc., hence may have lower costs for RES deployment. %Note that in Scotland or other countries such as Denmark, local groups often act together to make land available and invest in RES projects. In Scotland, Community Energy Scotland (CES) is an umbrella organisation of such groups and DNOs have an incentive to work with them.

Building the line will elicit a reaction from local investors. Crucially, the line investor has a \emph{first mover} advantage in building the grid infrastructure, which is expensive, technically challenging, and only few investors (e.g. DNO-approved) have the expertise and regulatory approval to carry it out. From a game-theoretic perspective, this is a bi-level Stackelberg game, between the \textit{leader} (line investor or first player) and \textit{follower} (local generators or second player).%\cite{von2010market}.
~We assume that players' actions are driven by profit maximisation criteria. The profit functions of players can be expressed as in~\cite{andoni2017game}:
\begin{equation}
\Pi_1=(E_{G_1}-E_{C_1})p_G-E_{G_1}c_{G_1}+(E_{G_2}-E_{C_2})p_T-C_T
\label{Prof_1}
\end{equation}
\begin{equation}
\Pi_2=(E_{G_2}-E_{C_2})(p_G-p_T)-E_{G_2}c_{G_2}
\label{Prof_2}
\end{equation}
In these equations, $\textstyle {E_{G_i}}$ represents player's $i$ expected energy produced over the project lifetime, if no curtailment occurred, while $\textstyle {E_{C_i}}$ is the energy lost through curtailment. Line cost is estimated as $\textstyle {C_T=I_T+M_T}$, where $\textstyle {I_T}$ is the cost of building the line (or initial investment) and $\textstyle {M_T}$ the cost of operation and maintenance. The monetary value of the power line is proportional to the energy flowing from B to A, charged for local generators and `common access' rules with $\textstyle {p_T}$ transmission fee per energy unit transported through the line. %Moreover, for a generation unit $i$, 
~Moreover, the cost of expected generation per unit is defined as $\textstyle {c_{G_i}=(I_{G,i}+M_{G_i})/{E_{G_i}}}$, where $\textstyle {I_{G_i}}$ is the cost of building the plant and $\textstyle {M_{G_i}}$ the operation and maintenance costs. Finally, the energy generated by a RES unit is sold at a constant generation tariff price, equal to $\textstyle {p_G}$.

As seen in Eq.~(\ref{Prof_1}), the line investor has two streams of revenue, the self-produced energy and the energy produced by local generators and transported through the line. The costs of the line investor relate to installing and operating the RES capacity ($c_{G_1}$) and to building the power line ($C_T$). Similarly from Eq.~(\ref{Prof_2}), the profits of the local generators depend only on the energy produced (generation cost $c_{G_2}$) and  transmitted through the line at a charge of $p_T$.

The research question our model tries to answer is \textit{`How to determine the optimal generation capacities $P_{N_i}$ built by the two players, so that profits are maximised?'}
~To answer this question, the profit equations, Eq.~(\ref{Prof_1}) and  Eq.~(\ref{Prof_2}), need to be expressed in terms of generation capacities.

Following the analysis presented in~\cite{andoni2017game}, expected generation $\textstyle {E_{G_i}}$ and curtailment $\textstyle {E_{C_i}}$ can be expressed as functions of the players' generation capacities. If $x_i$ is the per unit power generated by $i$ player, then this represents a stochastic variable that depends on the wind speed distribution, and is equal to $x_i={P_{G_i}}/{P_{N_i}}$, where $P_{G_i}$ is the actual power output of $i$ generator and $P_{N_i}$ its rated capacity. If $\mathbb{E}(P_{G_{i,t}})$ is the expected power generated at time interval $t$, then the total energy generated for duration equal to the project's lifetime is~$\displaystyle{E_{G_i}=\sum_t x_iP_{N_i}=\sum_t \mathbb{E}(P_{G_{i,t}}), \forall t}$. Similarly, the total energy curtailed is $\displaystyle{E_{C}=\sum_t \mathbb{E}(P_{C_{t}}), \forall t}$, where $\displaystyle{\mathbb{E}(P_{C_{t}})}$ is the expected power curtailed at time step $t$.

The resulting curtailment depends on wind resources at location B and demand at A, denoted as $P_{D,t}$. Curtailment events happen when $x_1P_{N_1}+x_2P_{N_2}-P_{D,t}>0$. Players' generating plants are located at neighbouring locations at B, therefore experience correlated wind speeds. The stochastic variables $x_1$ and $x_2$, follow a joint probability distribution function  $f(x_1,x_2)$ and expected curtailment at time interval $t$ can be expressed as (detailed analysis shown in~\cite{andoni2017game}):
\begin{equation}
\begin{aligned}
\mathbb{E}({P_{C,t}})= & \int_0^1\int_{\frac{P_{D,t}-x_1P_{N_1}}{P_{N_2}}}^1(x_1P_{N_1}+x_2P_{N_2})f(x_1,x_2)dx_2dx_1 \\ & - P_{D,t}\int_0^1\int_{\frac{P_{D,t}-x_1P_{N_1}}{P_{N_2}}}^1f(x_1,x_2)dx_2dx_1
\end{aligned}
\label{Tot_C}
\end{equation}
%Eq.~(\ref{Tot_C}) represents the expected power curtailed at time step $t$. If a longer time period is considered (e.g. equal to the project lifetime), the total energy curtailed is $\displaystyle{E_{C}=\sum_t E(P_{C_{t}}), \forall t}$.
Crucially, Eq.~(\ref{Tot_C}) shows that curtailment depends on both players' strategies i.e. the generation capacities built. Curtailment expressions for each player under a `common access' regime can be reasonably can be approximated by $E_{C_i}=\displaystyle{\frac{E_{G_i}}{E_{G_i}+E_{G_{-i}}}E_C}$ .%where $-i$ denotes the other player..
~This concludes the expression of profit equations as functions of players' rated capacities.

Optimal capacities installed are determined in the equilibrium of the game, which is found by \emph{backward induction}. The line investor or leader assesses and evaluates the second player's reaction, in order to determine his strategy (i.e. $\textstyle {P_{N_1}}$) and influence the equilibrium price. The leader estimates the follower's \emph{best response}, given his own capacity $P_{N_1}$:%, and then decides a strategy that maximises their profit. Given the generation capacity installed by the line investor $P_{N_1}$, the local generators best response is:
\begin{equation}
P_{N_2}^{*}= \argmax_{P_{N_2}} {\Pi_2(P_{N_1},P_{N_2}})
\label{BestResp1}
\end{equation}
Next, the leader estimates which solution from the set of the local generators' best response $P_{N_2}^{*}$ maximises his own profit:% Given the capacity built by the followers $P_{N_2}^{*}$, the line investor's best response is:
\begin{equation}
P_{N_1}^{*}= \argmax_{P_{N_1}}\Pi_1(P_{N_1},P_{N_2}^{*})
\label{BestResp2}
\end{equation}
%Local generators or \emph{followers} can only act after observing the leader's strategy. Note here that the terms line investor and leader (player 1) are used interchangeably, as are local generators and follower (player 2).  
%At a second stage, the follower observes this strategy and decides on their generation capacity to be installed, given by the best response function, i.e. maximising their own profit, as anticipated by the leader. 
In other words, the leader moves first by installing their own capacity. In the second level, followers respond to the capacity built, as anticipated by the leader. The equilibrium of the game  $(P_{N_1}^*,P_{N_2}^*)$ satisfies both Eq.~(\ref{BestResp1}) and Eq.~(\ref{BestResp2}) and is given by the notion of the subgame perfect equilibrium.

%under the adopted Principle of Access. For simplicity, we assume there is no RES capacity installed at location $\textstyle {B}$ prior to the construction of the power line.

In practical settings the joint distribution of stochastic renewable resources is often unknown, but historic data may be available. In addition, due to the interdependency in resulting curtailment and multiple parameters a nice closed-form solution of the game cannot be found or expressed analytically. In~\cite{andoni2017game} we presented an empirical algorithm that utilises directly real data and approximates the solution of the game following a one-shot approach. In addition, data may experience important gaps. In this paper we show how we can utilise real data to simulate scenarios that approximate the real distribution with a state-of-the-art MCMC technique.

\section{Gibbs sampling}
Markov chain Monte Carlo (MCMC) is a class of methods for simulation of stochastic processes. Gibbs sampling can be thought of as a particular case of the Metropolis-Hastings algorithm used for MCMC~\cite{fox2014convergence}. The Gibbs sampler uses the conditional distributions as proposal distributions with acceptance probability equal to 1~\cite{geyer1992practical} and can be easily implemented in various applications.%when conditional distributions are available. In general, iterative samplers are an attractive option when drawing samples from high dimensional multivariate distributions due to their inexpensive cost per iteration and small computer memory requirements~\cite{fox2014convergence}.
~Using this technique we can generate, from historic data, observations that are dependent. Wind data samples from the project locations form a Markov chain (MC). We can experiment with the length of the chain or \textit{sampling size} $n$ and we can repeat the process for multiple MCs or \textit{number of realisations} $N$. In practice, $n$ and $N$ need to be determined in such way that the resulting MC converges to the real distribution, is ergodic and computationally efficient. Ergodicity means that all possible states of the MC can be visited and are independent of the starting state~\cite{geyer1992practical}. The methodology applied is described in detail in the next sections with the help of the Kintyre-Hunterston case study.

%Gibbs sampling it allows us to generate observations from the conditional distribution using available historic data. A MCMC is a class of methods in which one can simulate draws that are slightly dependent and approximately from the posterior distribution~\cite{gupta2017classical}.

%In practice to implement Gibbs sampler we have to make sure that the resulting Markov Chain converges into the real distribution and this can happen independent of the starting state. We also need to make sure that the resulting MC is ergodic, or in other words that all possible states can be visited~\cite{geyer1992practical}. The satisfaction of these criteria depends on the sampling size $n$ and the number of realisations of the experiments $N$. Longer and multiple runs are possible however this is computationally expensive and therefore the right balance must be found.

%We show the Gibbs sampling methodology and equilibrium estimation in a concrete case study of a real grid reinforcement project in the UK, described in the following section.

\section{Case study analysis}

\subsection{Kintyre-Hunterston link}

\begin{figure}
\centering
\includegraphics[width=0.9\columnwidth]{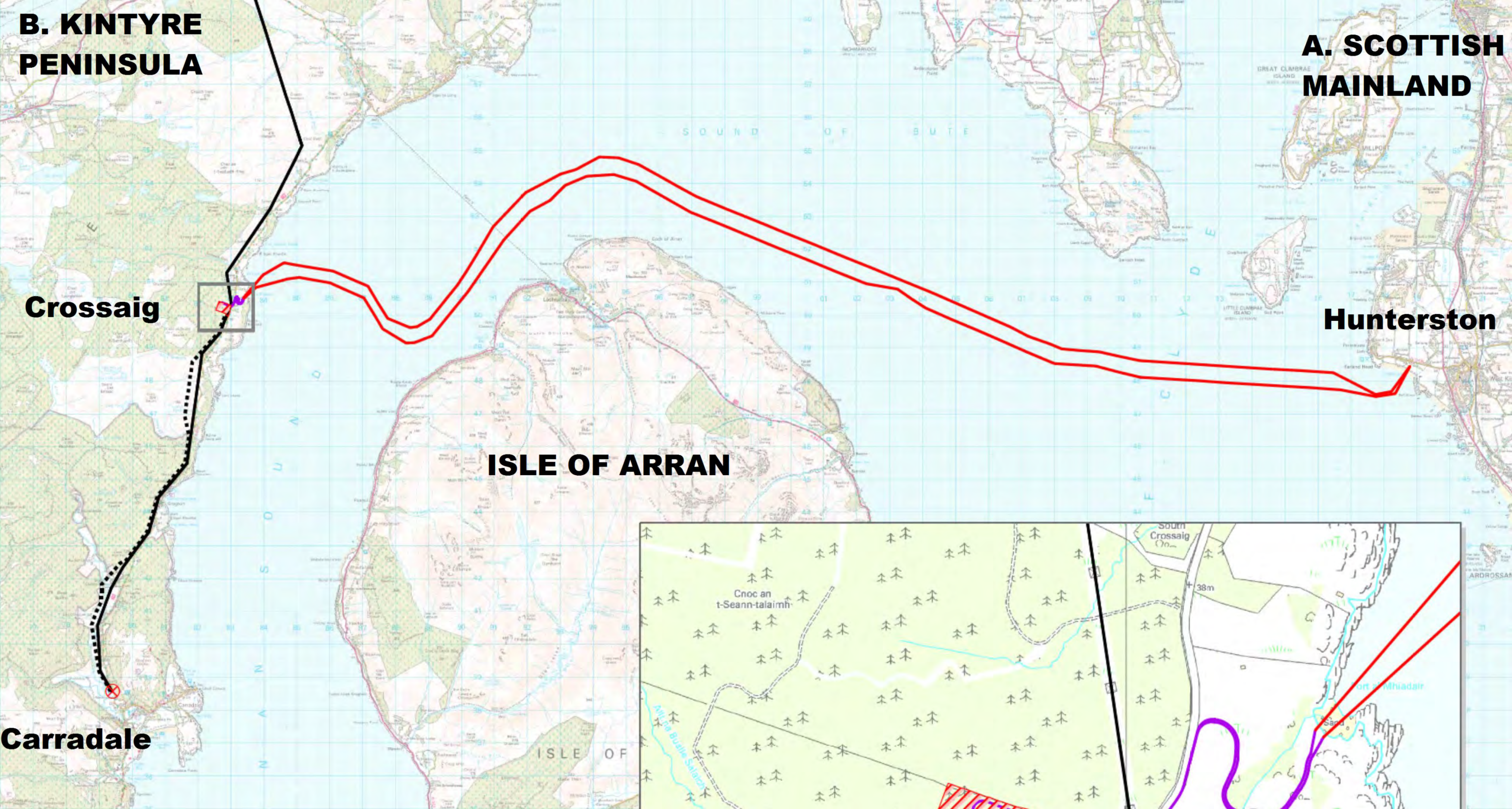}
\caption[Caption]{Kintyre-Hunterston project map\protect \footnotemark}
\label{fig:Kint-Hunt}
\end{figure}

The case study analysis is based on a real $\pounds 230\rm{m}$ grid reinforcement project in the UK that links the Kintyre peninsula to the Hunterston substation on the Scottish mainland (see Fig.~\ref{fig:Kint-Hunt}).%The project includes the installation of new overhead power lines, a new substation and a double circuit subsea cable of 220kV HVAC placed north of Arran for a distance of 41 km.
~Kintyre is a region that has attracted vast renewable investment, predominantly wind generation, resulting in the necessity of a newly built power line, which provided space for $150~\text{MW}$ of additional renewable capacity.% at a $\pounds 230\rm{m}$ cost.% In fact, the growth of renewable generation in the Kintyre region was responsible for the growing stress in the existing transmission line, originally designed and built to serve a typical rural area of low demand. According to SSE (the DNO in this region), the Kintyre-Hunterston project will provide $150~\text{MW}$ of additional renewable capacity and it will cost $\pounds 230\rm{m}$. Apart from facilitating renewable generation, the project is expected to increase security of supply and export capability to the mainland grid, delivering value to consumers estimated at $\pounds 18\rm{m}$ per annum \cite{SKM04}.

\begin{table*}[t]
  \centering
  \caption{Results for \protect $N=100$ realisations and an increasing number of observations \protect $n$}
  \begin{adjustbox}{max width=\textwidth}
\begin{tabular}{lcccccccccccccc}
 \hline
Sample size & $\bar w_1$  & ~$\sigma_{\bar w_1}$ & WCI $\bar w_1$ & ME $\bar w_1$ & $\bar w_2$  & ~$\sigma_{\bar w_2}$ & WCI $\bar w_2$ & ME $\bar w_2$ & $\bar P_D$  & ~$\sigma_{\bar P_D}$ & ~WCI $\bar P_D$ & ME $\bar P_D$\\
 \hline
$n=1,000$ & $12.1321$  & $0.6569$ & $0.2607$ & $18.91\%
$ & $12.2297$  & $0.6226$ & $0.2470$ & $17.61\%$ & $108.5722$  & $0.7859$ & $0.3119$ & $2.63\%$\\

$n=5,000$ & $12.0853$  & $0.2903$ & $0.1152$ & $6.65\%
$ & $12.1762$  & $0.2797$ & $0.1110$ & $6.47\%$ & $108.6271$  & $0.3395$ & $0.1348$ & $1.49\%$\\

$n=10,000$ & $12.0929$  & $0.2262$ & $0.0897$ & $4.63\%
$ & $12.1842$  & $0.2187$ & $0.0869$ & $4.36\%$ & $108.6000$  & $0.2403$ & $0.0953$ & $1.03\%$\\

$n=50,000$ & $12.1125$  & $0.0874$ & $0.0347$ & $2.02\%
$ & $12.2028$  & $0.0857$ & $0.0340$ & $1.97\%$ & $108.5979$  & $0.1033$ & $0.0410$ & $0.73\%$\\

$n=100,000$ & $12.1155$  & $0.0631$ & $0.0251$ & $1.28\%
$ & $12.2075$  & $0.0602$ & $0.0239$ & $1.16\%$ & $108.5954$  & $0.0663$ & $0.0263$ & $0.70\%$\\

$n=200,000$ & $12.1065$  & $0.0441$ & $0.0175$ & $0.80\%
$ & $12.1986$  & $0.0427$ & $0.0169$ & $0.76\%$ & $108.5915$  & $0.0453$ & $0.0180$ & $0.62\%$\\

$n=500,000$ & $12.1049$  & $0.0272$ & $0.0108$ & $0.68\%
$ & $12.1968$  & $0.0262$ & $0.0103$ & $0.62\%$ & $108.5930$  & $0.0288$ & $0.0114$ & $0.57\%$\\
 \hline
\end{tabular}
\end{adjustbox}
\label{Tab:sample}
\end{table*}

\begin{table*}[t]
  \centering
  \caption{Results for \protect $n=5,000$ sampling size and an increasing number of realisations \protect $N$}
  \begin{adjustbox}{max width=\textwidth}
\begin{tabular}{lcccccccccccccc}
 \hline
Realisations & $\bar w_1$  & ~$\sigma_{\bar w_1}$ & WCI $\bar w_1$ & ME $\bar w_1$ & $\bar w_2$  & ~$\sigma_{\bar w_2}$ & WCI $\bar w_2$ & ME $\bar w_2$ & $\bar P_D$  & ~$\sigma_{\bar P_D}$ & ~WCI $\bar P_D$ & ME $\bar P_D$\\
 \hline
$N=100$ & $12.0850$  & $0.2774$ & $0.0840$ & $6.64\%
$ & $12.2017$  & $0.2671$ & $0.0809$ & $6.01\%$ & $108.5703$  & $0.3037$ & $0.0919$ & $1.34\%$\\

$N=170$ & $12.1120$  & $0.2217$ & $0.0879$ & $5.35\%
$ & $12.1812$  & $0.2145$ & $0.0851$ & $5.15\%$ & $108.6275$  & $0.3279$ & $0.1301$ & $1.26\%$\\

$N=500$ & $12.0867$  & $0.2709$ & $0.0476$ & $7.22\%
$ & $12.1764$  & $0.2618$ & $0.0460$ & $7.16\%$ & $108.5932$  & $0.3052$ & $0.0536$ & $1.47\%$\\

$N=1,000$ & $12.1097$  & $0.2764$ & $0.0343$ & $8.20\%
$ & $12.2025$  & $0.2666$ & $0.0331$ & $7.57\%$ & $108.5949$  & $0.3198$ & $0.0397$ & $1.47\%$\\

$N=5,000$ & $12.1044$  & $0.2793$ & $0.0155$ & $9.06\%
$ & $12.1966$  & $0.2717$ & $0.0151$ & $8.81\%$ & $108.5951$  & $0.3244$ & $0.0180$ & $1.58\%$\\

$N=10,000$ & $12.1032$  & $0.2754$ & $0.0108$ & $9.41\%
$ & $12.1956$  & $0.2672$ & $0.0105$ & $8.60\%$ & $108.5917$  & $0.3268$ & $0.0128$ & $1.70\%$\\

$N=50,000$ & $12.1022$  & $0.2787$ & $0.0049$ & $9.87\%
$ & $12.1943$  & $0.2707$ & $0.0048$ & $9.26\%$ & $108.5901$  & $0.3241$ & $0.0057$ & $1.70\%$\\
 \hline
\end{tabular}
\end{adjustbox}
\label{Tab:real}
\end{table*}

Following the analysis described in Section~\ref{Gen_model}, we assume that the demand region or Location A is Hunterston and location B is the geographical region covering the Kintyre peninsula. The line investor and local generators install generation capacities at different sub-regions of B.% and local generators install wind capacity at a different sub-region of location B. 
~Two weather stations were selected by the UK Met Office database\footnote{\url{https://badc.nerc.ac.uk/search/midas_stations/}}, the station with ID 908 located in the Kintyre peninsula (wind farm of line investor) and with ID 23417 located in Islay (wind farm of local generators), with a distance between them of $44~\rm{km}$. These weather stations provide data over a common period of 17 years (1999--2015). Demand data used in simulations are based on real UK National Demand data~\footnote{\url{http://www2.nationalgrid.com/UK/Industry-information/Electricity-transmission-operational-data/Data-Explorer/}} in the time period of 2006--2015. UK demand data are normalised to represent a lower local demand. More details on the case study and data processing can be found in~\cite{andoni2017game}.

Literature in wind forecasting%~\cite{fruh2015local} 
~commonly uses Weibull distributions for the representation of actual wind distributions. However, the joint probability distribution of the wind speed (and of the players' power outputs), exhibits correlation and in practice is not known. If there are sufficient wind speed measurements for both players locations, then the joint probability distribution can be approximated directly from the available historic data. The method described in the following section can be used to draw observations from available data and simulate different scenarios. The technique can generate large datasets as required for the intended analysis.

\subsection{Gibbs sampling applied}

\begin{algorithm}
\caption{\textsc{Gibbs Sampling}}\label{gibbs_sampling}
\begin{algorithmic}[1]
\State $w_1$,$w_2$,$P_{D}$ \Comment{wind speed 1,2, power demand}
\State $T$ \Comment{number of samples}
\State $\langle w_1^{(k)},w_2^{(k)},P_D^{(k)} \rangle$, $k\in  \{1,2,...,k_{max}\}$ \Comment{historic data}
\State $F(w_1,w_2)$ \Comment{wind distribution from data}
\State $G\left(P_D,\displaystyle{\frac{w_1+w_2}{2}}\right)$ \Comment{demand cond. distrib. on mean wind}
\State $t \gets 1$
\State $\langle w_1^{(t)},w_2^{(t)},P_D^{(t)} \rangle \gets sample(w_1,w_2,P_D)$ \Comment{initialise}
\State \textbf{repeat}
\Indent
\State $w_1^{(t+1)} \gets sample$ $F(w_1\mid _{w_2^{(t)}})$
\State $w_2^{(t+1)} \gets sample$ $F(w_2\mid _{w_1^{(t+1)}})$
\State $P_D^{(t+1)} \gets sample$ $G\left( P_D\mid _{\textstyle{\frac{w_1^{(t+1)}+w_2^{(t+1)}}{2}}}\right)$
\State $t \gets t+1$
\EndIndent
\State \textbf{until} $t>T$
\State \textbf{return} $\langle w_1^{(t)},w_2^{(t)},P_D^{(t)} \rangle$, $t\in  \{t_{burn},t_{burn}+1,...,T\}$
\end{algorithmic}
\end{algorithm}
\label{Alg:Gibbs}

We apply Gibbs sampling to  the joint bivariate distribution of wind speeds at the players' locations. Players' wind speeds at each time $t$ form a MC. The methodology is described in Alg.~1. From available historic data, we create a joint distribution table of wind speeds at the players' locations. For every possible wind speed $w_1$ of first player, we record the subset of $w_2$ wind speed, and vice versa (Line 4 in Alg.~1). This represents the conditional distributions. In practice,the probabilities for certain combinations of wind speeds can be low (e.g. it is unlikely to have extremely high wind at one location and low wind speed to a proximal location), therefore some subsets can be sparse, either because some of observations represent rare events or due to correlation. We overcome this difficulty by merging sparse bins of rare events and outliers, and ensure ergodicity of the MC. The MC is initialised by randomly selecting a sample from the joint distribution table (Line 7 in Alg.~1). Each iteration step involves replacing the value of one variable by a value selected randomly by the conditional $F(w_i\mid _{w_-i^{(t)}})$. In addition, demand is randomly selected by the conditional distribution of demand over the average wind speed (Line 11 in Alg.~1). The procedure is cycled through the variables forming $n$ samples of $\langle w_1^{(t)},w_2^{(t)},P_D^{(t)} \rangle$, $t\in  \{1,2,...,T=n\}$. To ensure that the MC converges, we run Alg.~1 for several sampling sizes $n$ (small, moderate, large) and repeat the procedure for $N$ realisations. Results are shown in Tables~\ref{Tab:sample} and~\ref{Tab:real}. Columns represent the sample mean, standard deviation of the sample mean, width of the 95\% confidence interval (WCI) and maximum error (ME) from mean of historic data, i.e.  $\mu_{w_1}=12.1029$, $\mu_{w_1}=12.1950$ and $\mu_{P_D}=108.1830$. As the sampling size increases the sample mean follows a normal distribution and the standard deviation decreases, as expected by the central limit theorem (Table~\ref{Tab:sample}). We perform the same analysis for an increasing number of realisations that use a different starting point (Table~\ref{Tab:real}). We also adopt a burn-in or warm-up period of 20\% of samples to make sure that our results are independent off the starting state~\cite{geyer1992practical}. The results show that MC converges to the distribution from data and that a large $n$ is required but $N$ can be chosen to be relatively small. For this reason and driven by computational limitations, for the estimation of the Stackelberg equilibrium, we selected $n=50,000$, $N=170$ and burn-in of $10,000$ samples.

%Based on~\cite{geyer1992practical} we try to find a balance between multiple shorter runs and one-shot longer runs. We have some prior knowledge of the joint distribution. For example, a single wind speed can be approximated by a Weibull distribution. Independent wind speeds could also be approximated by a bivariate Weibull distribution however in our occasion the two distributions experience some level of correlation.

\subsection{Stackelberg equilibrium estimation}
Wind speed data generated by the Gibbs sampler are used to estimate the per unit power output of wind generators. Estimation is based on a generic power curve\footnote{Parameters derived by a $2.05~\text{MW}$ Enercon E82 wind turbine: \url{http://www.enercon.de/en/products/ep-2/e-82/}} and a sigmoid function approximation (see Alg.~2 Line 4). Players' strategies are the capacities they can install. The maximum feasible solution for a single player was set equal to $P_{N_{max}}=500.5~\text{MW}$ and the incremental  capacity to $step=0.5~\text{MW}$. For every possible combination of the rated capacities installed $(P_{N_1},P_{N_2})$, we estimate the power generated and curtailed for each player on an hourly basis. Next, we estimate the aggregate power generated and curtailed by each player as the summation of $40,000$ data points. The procedure is described in Alg.~2.

\begin{algorithm}[t]
\caption{\textsc{Generation \& Curtailment Estimation}}\label{energy_estimation}
\begin{algorithmic}[1]
\State $P_{N_{max}}$ \Comment{max rated capacity in search space}
\State $P_{N_i} \gets [0:step:P_{N_{max}}]$ \Comment{strategy space i=1,2 player}
\State $\alpha$, $\beta$  \Comment{power curve sigmoid parameters}
\State $P_{G_i}^{(t)} \gets \displaystyle{\frac{1}{1+e^{-\alpha(w_1^{(t)}-\beta)}}}\cdot P_{N_i}$ \Comment{generation i player}
\State \textbf{for all} $\textit{P}_{N_1}\in \{0,...,P_{Nmax}\}$ \textbf{do}
\Indent
\State \textbf{for all} $\textit{P}_{N_2}\in \{0,...,P_{Nmax}\}$ \textbf{do}
\Indent
\State $RD \gets P_D-(P_{G_1}+P_{G_2})$ \Comment{residual demand}
\If {$\textit{RD} > 0$} \Comment{no curtailment}
\State $RD \gets 0$
\EndIf
\State \textbf{end}
\State $P_{C_1} \gets \displaystyle{\frac{P_{G_1}\cdot RD}{P_{G_1}+P_{G_2}}}$ \Comment{curtailment gen 1}
\State $P_{C_2} \gets \displaystyle{\frac{P_{G_2}\cdot RD}{P_{G_1}+P_{G_2}}}$ \Comment{curtailment gen 2}
\EndIndent
\State \textbf{end}
\EndIndent
\State \textbf{end}
\State $E_{G_i}(P_{N_i}) \gets \displaystyle{\sum P_{G_i}}$ \Comment{total gen i}
\State $E_{C_1}(P_{N_1},P_{N_2}) \gets \displaystyle{\sum P_{C_1}}$ \Comment{total curt 1}
\State $E_{C_2}(P_{N_1},P_{N_2}) \gets \displaystyle{\sum P_{C_2}}$ \Comment{total curt 2}
\State \textbf{return} $(E_{G_1}, E_{G_2}, E_{C_1}, E_{C_2})$
\end{algorithmic}
\end{algorithm}

For several cost parameters $(c_{G_1},c_{G_2},p_T)$ and feed-in tariff price $p_G$, we estimate the profits as defined in Eq.~(\ref{Prof_1}) and Eq.~(\ref{Prof_2}). For every possible $P_{N_1}$, we find the capacity $P_{N_2}^*$ that maximises the follower's profits $\Pi_2^*$ (follower's best response). From this set of solutions, the leader selects the one that maximises its own profit i.e. $P_{N_1}^*$ (leader's best response). The equilibrium of the game is given by the pair $(P_{N_1}^*,P_{N_2}^*)$, which satisfies best response functions as  described in Alg.~3.

\begin{algorithm}
\caption{\textsc{Profit \& Stackelberg Equilibria}}\label{profit_stackelberg}
\begin{algorithmic}[1]
\State $\textit{p}_{G}$,~$\textit{p}_{T}$ \Comment{feed-in tariff, transmission fee}
\State $\textit{C}_{T}$ \Comment{cost of line}
\State $\textit{c}_{G_i}$ \Comment{i player's generation cost}
\State \textbf{for all} $\textit{P}_{N_1}\in \{0,...,P_{N_{max}}\}$ \textbf{do}
\Indent
\State \textbf{for all} $\textit{P}_{N_2}\in \{0,...,P_{N_{max}}\}$ \textbf{do}
\Indent
\State $\Pi_1 \gets{}(E_{G_1}-E_{C_1})p_G-E_{G_1}c_{G_1}+$\\
 \hspace{1.95cm} $(E_{G_2}-E_{C_2})p_T-C_T$
\State $\Pi_2 \gets {} (E_{G_2}-E_{C_2})(p_G-p_T)-E_{G_2}c_{G_2}$
\EndIndent
\State \textbf{end}
\EndIndent
\State \textbf{end}
\State \textbf{for all} $\textit{P}_{N_1}\in \{0,...,P_{N_{max}}\}$ \textbf{do} \Comment{best response gen 2}
\Indent
\State $\Pi_{2}^{*} \gets \displaystyle{\max_{P_{N_2}} {\Pi_2(P_{N_1},P_{N_2}})}$
\State $P_{N_2}^{*} \gets \displaystyle{\argmax_{P_{N_2}} {\Pi_2(P_{N_1},P_{N_2}})}$
\EndIndent
\State \textbf{end}
\State $\Pi_{1}^{*} \gets \displaystyle{\max_{P_{N_1}} {\Pi_1(P_{N_1},P_{N_2}^{*}})}$ \Comment{best response gen 1}
\State $P_{N_1}^{*} \gets \displaystyle{\argmax_{P_{N_1}} {\Pi_1(P_{N_1},P_{N_2}^{*}})}$
\State \textbf{return} $\Pi_1^*, \Pi_2^*,P_{N_1}^*, P_{N_2}^*$
\end{algorithmic}
\end{algorithm}

\section{Results and discussion}
\begin{figure}
\centering
\includegraphics[width=\columnwidth]{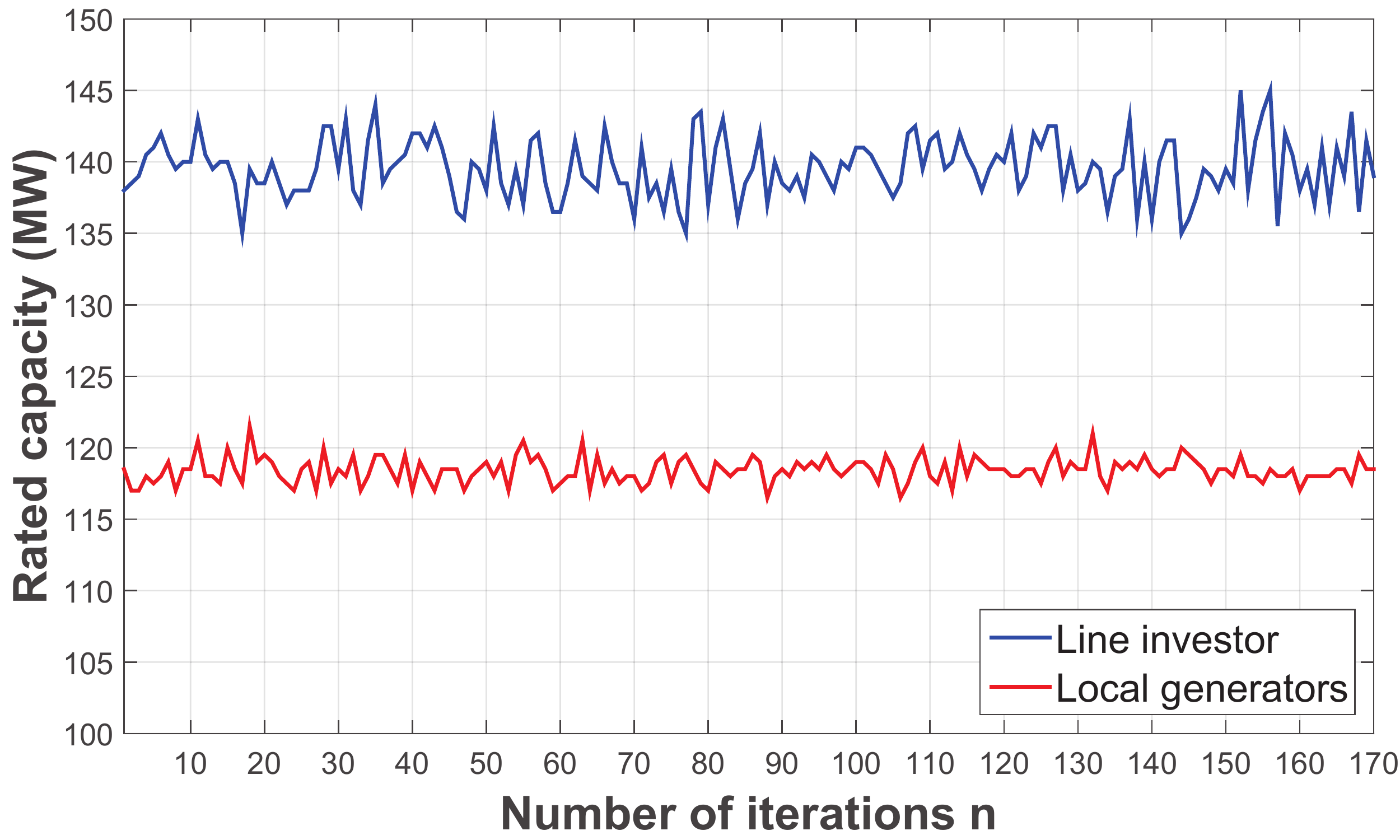}
\caption{Optimal generation capacities of each player for several realisations ($c_{G_1}=0.3p_G$, $c_{G_2}=0.28p_G$, $p_T=0.26p_G$ and $p_G=\pounds 74.3 /\text{MWh}$)}
\label{Scenario1_cap_iter}
\end{figure}

The methodology described in previous sections was followed to run several experiments. Fig.~\ref{Scenario1_cap_iter} shows the optimal rated capacities built by the players at the equilibrium of the game for $N=170$ realisations. Recall here that every realisation represents a completely different MC generated. The results are satisfactory and show a $10~MW$ range in the estimated solutions for optimal rated capacities. Similar results were observed for the optimal profits derived.
%First of all for a given set of cost parameters and $N=170$ realisations optimal rated capacities and profits of players are shown in Fig.~\ref{Scenario1_cap_iter} and Fig.~\ref{Scenario1_prof_iter}, respectively. Recall here that every realisation represents a completely different MC drawn by historic data. The results show a $10~MW$ range of estimated rated capacities.

\begin{comment}
\begin{figure}
\includegraphics[width=\columnwidth]{Scen1_Prof_iterat}
\caption{Optimal profits of each player for several realisations ($c_{G_1}=0.3p_G$, $c_{G_2}=0.28p_G$, $p_T=0.26p_G$ and $p_G=\pounds 74.3 /\text{MWh}$)}
\label{Scenario1_prof_iter}
\end{figure}
\end{comment}

\begin{figure*}
$\begin{array}{ccc}
\includegraphics[width=0.32\textwidth]{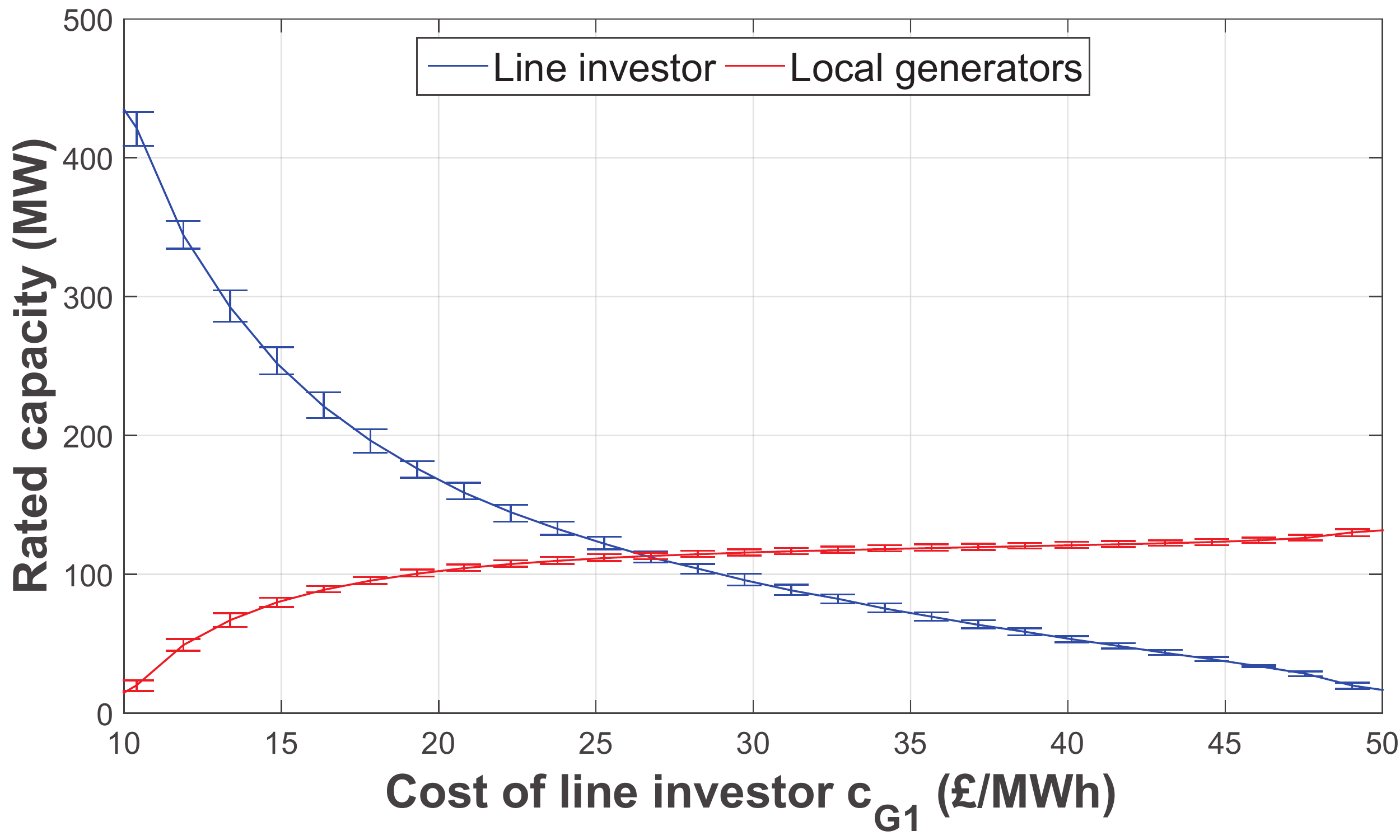} & \includegraphics[width=0.32\textwidth]{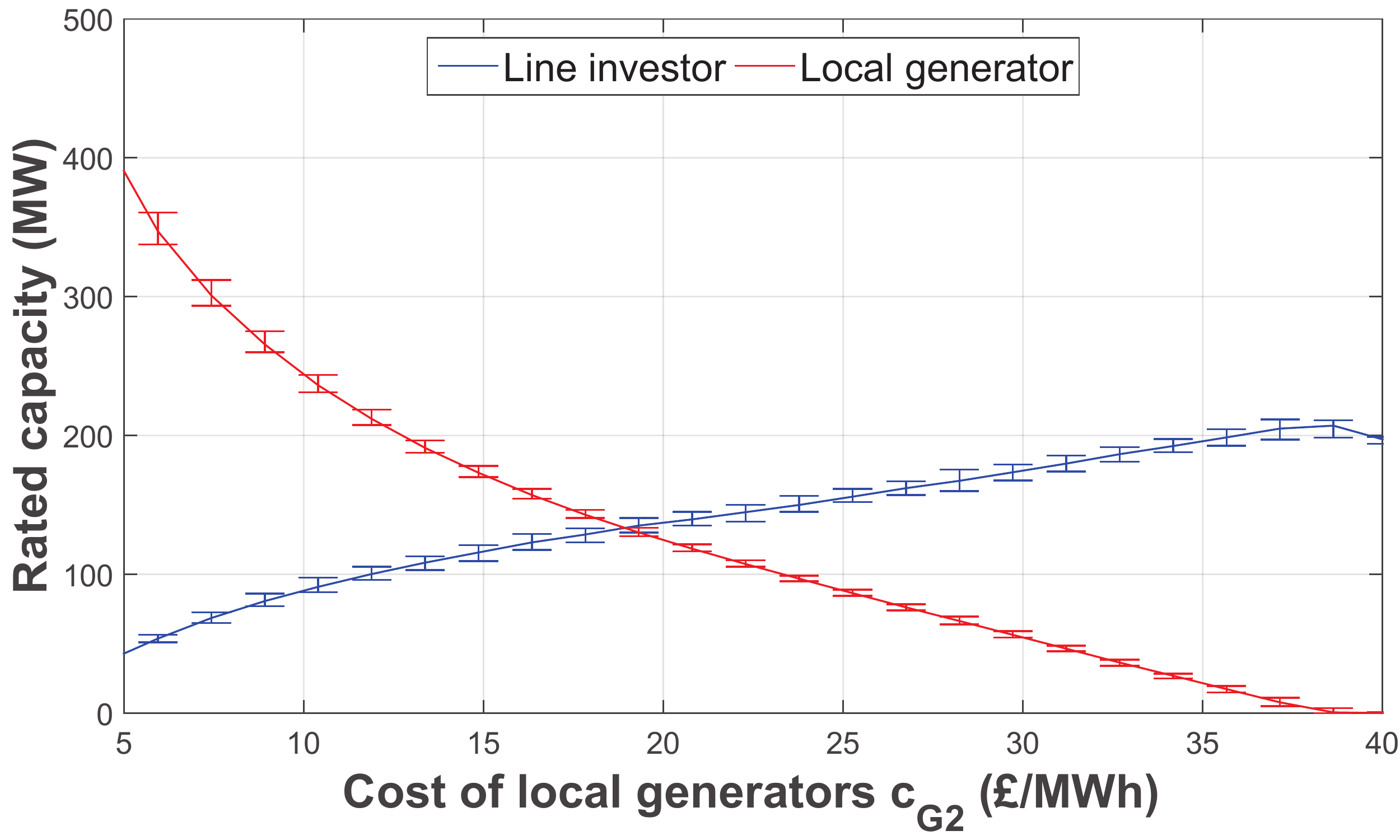} & \includegraphics[width=0.32\textwidth]{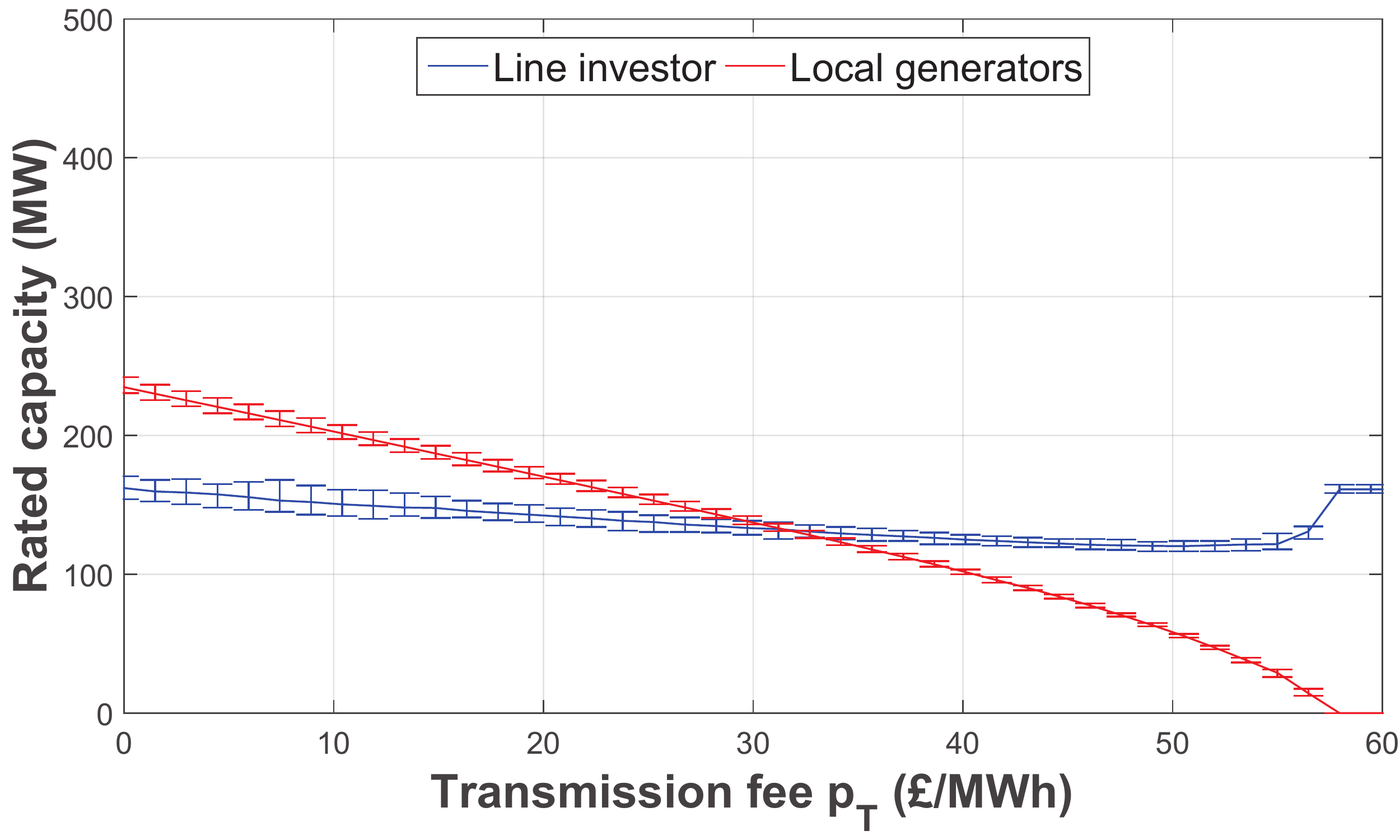} \\
\includegraphics[width=0.32\textwidth]{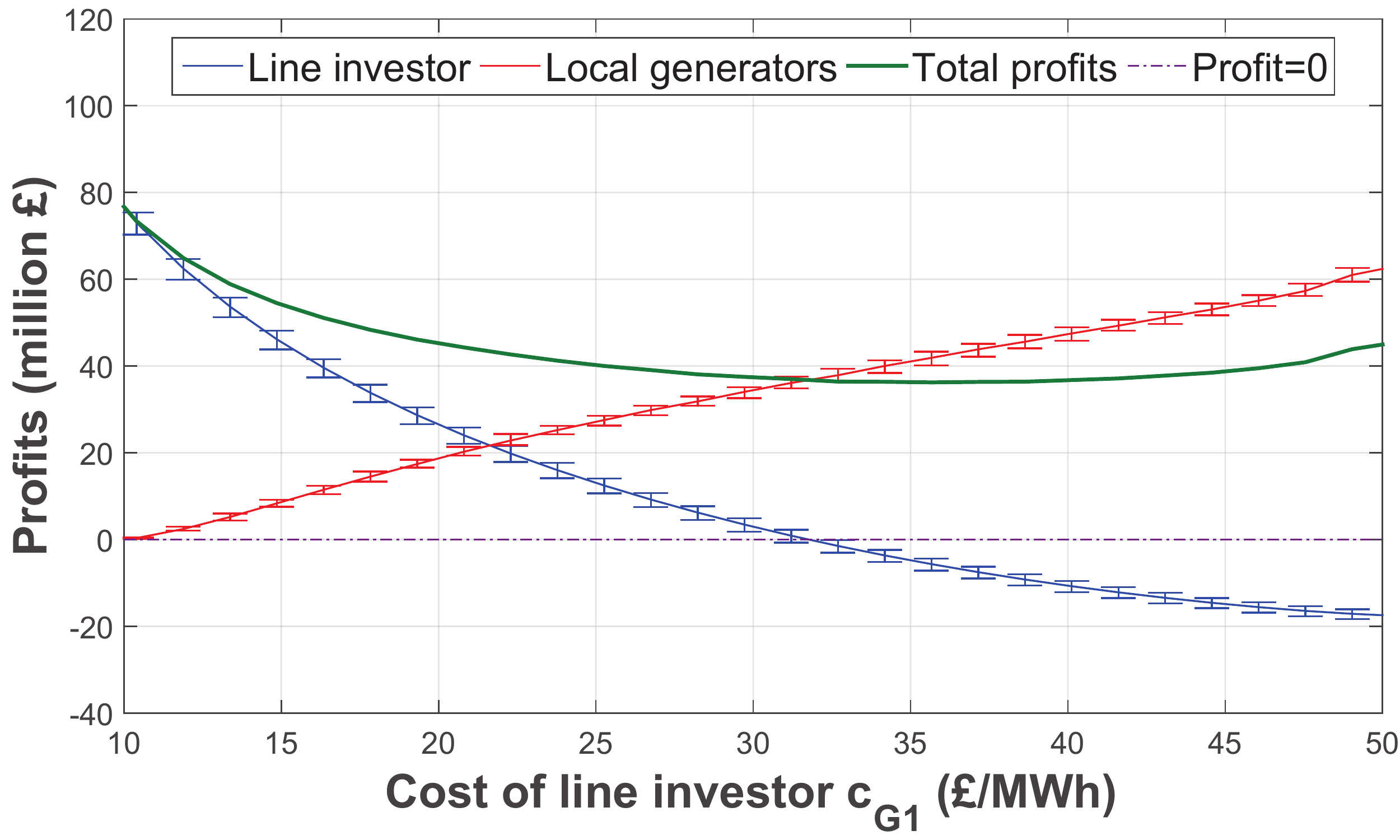} & \includegraphics[width=0.32\textwidth]{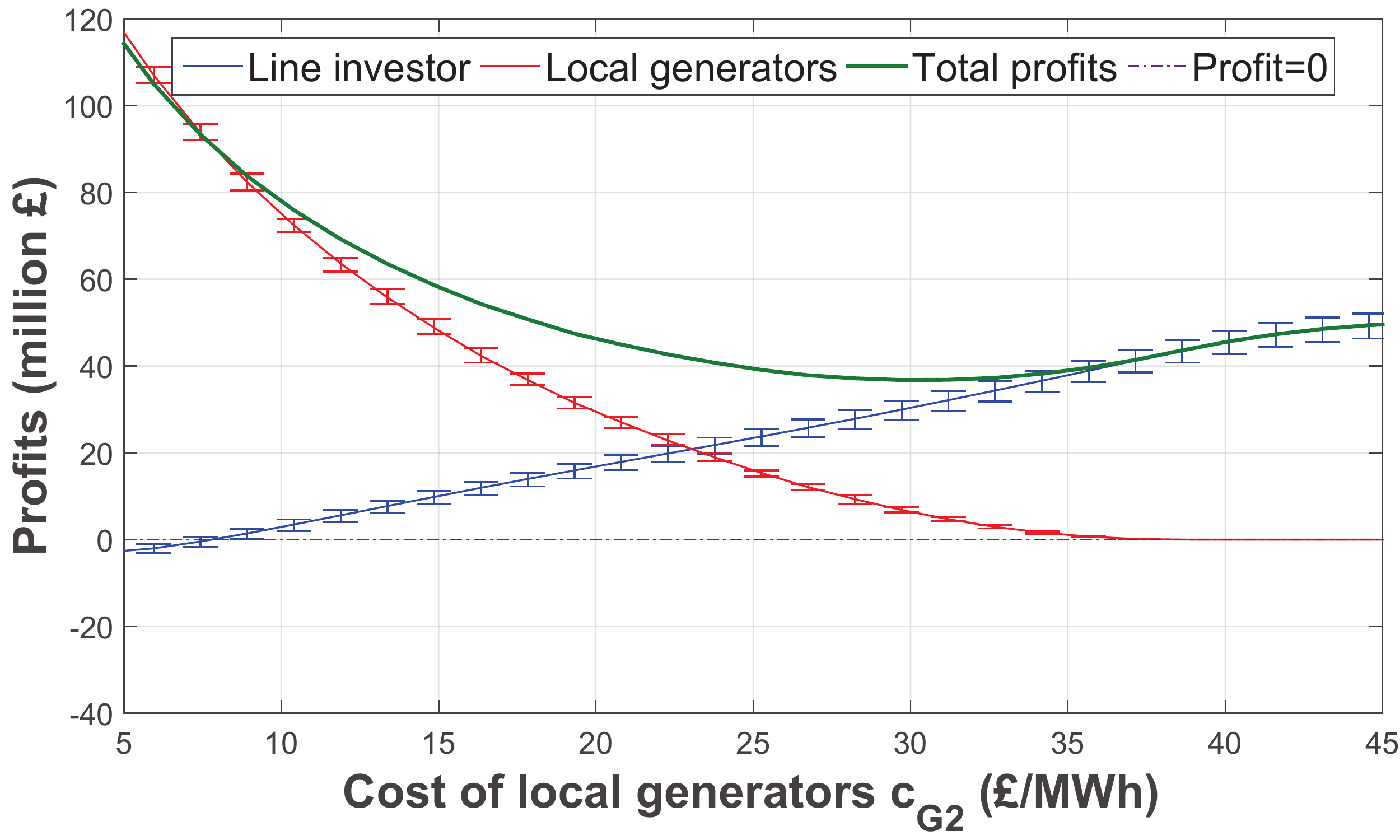} & \includegraphics[width=0.32\textwidth]{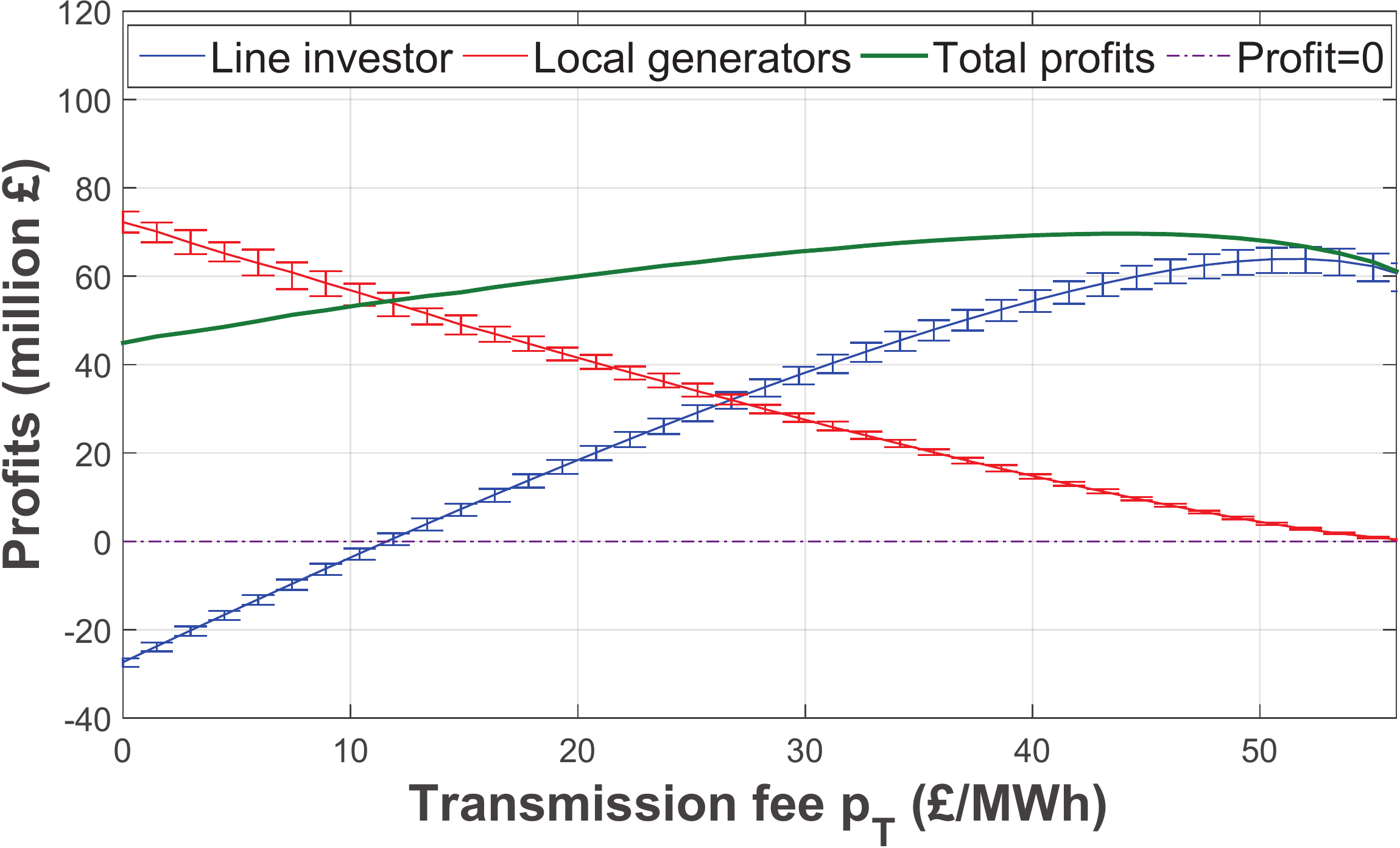}\\
\end{array}$
\caption{Rows (1) and (2) show generation capacity built and profits at Stackelberg equilibrium, respectively, column (1) shows dependency on generation cost of line investor, (2) on on generation cost of local generators and (3) on transmission fee}
\label{Stackel_graphs}
\end{figure*}

Moreover, we study how the equilibrium results depend on varying cost parameters. Fig.~\ref{Stackel_graphs} shows the dependence on line investor's cost (first column), local generators' cost (second column) and the transmission fee (third column). We assume that both players can sell the energy generated for $p_G=\pounds 74.3 /\text{MWh}$. For each scenario, the key parameter varies, while other parameters remain fixed (\emph{Scenario 1}: $c_{G_1}=0.16\ldots 0.68p_G$, $c_{G_2}=0.30p_G$ and $p_T=0.26p_G$, \emph{Scenario 2}: $c_{G_1}=0.30p_G$, $c_{G_2}=0.08\ldots 0.54p_G$ and $p_T=0.26p_G$ and \emph{Scenario 3}: $c_{G_1}=0.26p_G$, $c_{G_2}=0.20p_G$ and $p_T=0\ldots 0.80p_G$). The results in Fig.~\ref{Stackel_graphs} show the average equilibrium solution and min-max solutions, found for $N=170$ realisations of the simulation procedure.

%Recall here, the follower will install generation capacity $P_{N_2}$ as long as the revenues earned, depending on $p_G-p_T$, are larger than the cost of installing this capacity, depending on $c_{G_2}$. Crucially, the revenues depend on the curtailment imposed to the local generators $E_{C_2}$, which is interdependent on the capacity installed by the line investor $P_{N_1}$. On the other hand, the line investor earns revenues from the energy generated $E_{1}(P_{N_1})$, depending on $p_G$ and the energy transported through the line, $E_2(P_{N_2})$, which is charged with $p_T$. Both $E_1$ and $E_2$ depend on the curtailment imposed, which is a function of the rated capacities installed $E_C(P_{N_1},P_{N_2})$. The costs associated include the generation cost $c_{G_1}$ and the fixed cost of the line $C_T$. The line investor will install generation capacity himself as long as the cost of installing an additional generation unit results in increasing the profit.

In all sets of scenarios, the total capacity installed by all players decreases as the tested parameter value increases. Each player installs less capacity as their generation cost increases, while the other player benefits by increasing their capacity. The cost of local generators has a larger impact on the capacities installed for both players, as shown by comparing the first to the second column. Profits have similar behaviour to the optimal rated capacities, but local generators face the additional cost of transmission charges. If the followers' generation cost is much lower than the line investor's (assuming for example that local generators might have access to cheaper land), the line investor needs to charge a high transmission fee to have positive earnings. On the contrary, if the leader's cost is much lower, the generation capacity will mostly be installed by the line investor, as there is no room for profitable investment from local renewable producers. As shown in Scenario~3, $p_T\simeq0.16p_G$ or $\simeq \pounds 12\text{MWh}$ is the minimum value of transmission charges that allows profit for the line investor. Similarly, if the transmission fee is set too high, then it is not profitable for local investors to invest in renewable generation.%, as their profit diminishes with increasing transmission fee.
~As $p_T$ is set by the system regulator, the methodology can be useful to determine a feasible range of charges that allows both transmission and generation investments to be profitable.

\section{Conclusions \& Future Work}

In this work we show how privately developed network upgrade for DGs can lead to a leader-follower game between the line and local investors. Curtailment and line access rules play a key role in the strategic game, the equilibrium of which determines optimal generation capacities and their profits. %We use a profit maximising model that can offer good insights to the strategic game formed between players for varying cost parameters.
~Settings where this model can be applied include numerous locations where demand and generation are not co-located. When real historic data is available, we can use MCMC and Gibbs sampling to simulate multiple future scenarios and reduce the uncertainty of the investment decisions. In the future, we plan to extend the model to multi-location settings and introduce energy storage, which enables using renewable energy to satisfy more of the outstanding demand, and hence reduces curtailment, changing the joint investment game.
\vspace{-0.1cm}
\section*{Acknowledgements}
\vspace{-0.1cm}
The authors acknowledge the EPSRC National Centre for Energy Systems Integration (CESI) [EP/P001173/1].% and thank Community Energy Scotland, SSE, National Grid and the UK Met Office for data access and all information provided.

\bibliographystyle{IEEEtran}
\bibliography{mybiblio}

\begin{thebibliography}{10}
\providecommand{\url}[1]{#1}
\csname url@samestyle\endcsname
\providecommand{\newblock}{\relax}
\providecommand{\bibinfo}[2]{#2}
\providecommand{\BIBentrySTDinterwordspacing}{\spaceskip=0pt\relax}
\providecommand{\BIBentryALTinterwordstretchfactor}{4}
\providecommand{\BIBentryALTinterwordspacing}{\spaceskip=\fontdimen2\font plus
\BIBentryALTinterwordstretchfactor\fontdimen3\font minus
  \fontdimen4\font\relax}
\providecommand{\BIBforeignlanguage}[2]{{%
\expandafter\ifx\csname l@#1\endcsname\relax
\typeout{** WARNING: IEEEtran.bst: No hyphenation pattern has been}%
\typeout{** loaded for the language `#1'. Using the pattern for}%
\typeout{** the default language instead.}%
\else
\language=\csname l@#1\endcsname
\fi
#2}}
\providecommand{\BIBdecl}{\relax}
\BIBdecl

\bibitem{estopier2013flexible}
A.~{Laguna Estopier}, E.~{Crosthwaite Eyre}, S.~Georgiopoulos, and C.~Marantes,
  ``{FPP low carbon networks: Commercial solutions for active network
  management},'' in \emph{CIRED}, Stockholm, 2013.

\bibitem{kane2014review}
L.~Kane and G.~Ault, ``A review and analysis of renewable energy curtailment
  schemes and {Principles of Access}: Transitioning towards business as
  usual,'' \emph{Energy Policy}, vol.~72, pp. 67--77, May 2014.

\bibitem{andoni2016game}
M.~Andoni, V.~Robu, and W.-G. Fr{\"u}h, ``Game-theoretic modeling of
  curtailment rules and their effect on transmission line investments,'' in
  \emph{IEEE 2012 PES ISGT Europe}.\hskip 1em plus 0.5em minus 0.4em\relax
  IEEE, 2016, pp. 1--6.

\bibitem{michiorri2011dynamic}
A.~Michiorri, R.~Currie, P.~Taylor, F.~Watson, and D.~Macleman, ``{Dynamic line
  ratings deployment on the Orkney smart grid},'' in \emph{{CIRED}}, 2011.

\bibitem{anaya2015options}
K.~L. Anaya and M.~G. Pollitt, ``Options for allocating and releasing
  distribution system capacity: Deciding between interruptible connections and
  firm {DG} connections,'' \emph{Applied Energy}, vol. 144, pp. 96--105, 2015.

\bibitem{bronski2015economics}
P.~Bronski, J.~Creyts, M.~Crowdis, S.~Doig, J.~Glassmire, L.~Guccione
  \emph{et~al.}, \emph{The economics of load defection: How grid-connected
  solar-plus-battery systems will compete with traditional electric service-Why
  it matters, and possible paths forward}, 2015.

\bibitem{maurovich2015transmission}
L.~Maurovich-Horvat, T.~K. Boomsma, and A.~S. Siddiqui, ``Transmission and wind
  investment in a deregulated electricity industry,'' \emph{IEEE Trans. on
  Power Systems}, vol.~30, no.~3, pp. 1633--1643, May 2015.

\bibitem{perrault2014efficient}
A.~Perrault and C.~Boutilier, ``Efficient coordinated power distribution on
  private infrastructure,'' in \emph{Int. Conference on Autonomous Agents and
  Multi-Agent Systems}.\hskip 1em plus 0.5em minus 0.4em\relax Paris: {AAMAS},
  May 2014, pp. 805--812.

\bibitem{van2012economics}
A.~H. van~der Weijde and B.~F. Hobbs, ``The economics of planning electricity
  transmission to accommodate renewables: Using two-stage optimisation to
  evaluate flexibility and the cost of disregarding uncertainty,'' \emph{Energy
  Economics}, vol.~34, no.~6, pp. 2089--2101, Feb. 2012.

\bibitem{asimakopoulou2013leader}
G.~E. Asimakopoulou, A.~L. Dimeas, and N.~D. Hatziargyriou, ``Leader-follower
  strategies for energy management of multi-microgrids,'' \emph{IEEE
  Transactions on Smart Grid}, vol.~4, no.~4, pp. 1909--1916, Dec. 2013.

\bibitem{andoni2017game}
M.~Andoni, V.~Robu, W.-G. Fr{\"u}h, and D.~Flynn, ``Game-theoretic modeling of
  curtailment rules and network investments with distributed generation,''
  \emph{Applied Energy}, vol. 201, pp. 174--187, 2017.

\bibitem{papaefthymiou2008mcmc}
G.~Papaefthymiou and B.~Klockl, ``{MCMC for wind power simulation},''
  \emph{IEEE Trans. on Energy Conversion}, vol.~23, no.~1, pp. 234--240, 2008.

\bibitem{wu2012markov}
T.~Wu, X.~Ai, W.~Lin, J.~Wen, and L.~Weihua, ``{Markov chain Monte Carlo method
  for the modeling of wind power time series},'' in \emph{2012 IEEE PES ISGT
  Asia}.\hskip 1em plus 0.5em minus 0.4em\relax IEEE, 2012, pp. 1--6.

\bibitem{fox2014convergence}
C.~Fox and A.~Parker, ``{Convergence in variance of Chebyshev accelerated Gibbs
  samplers},'' \emph{SIAM Journal on Scientific Computing}, vol.~36, no.~1, pp.
  A124--A147, 2014.

\bibitem{geyer1992practical}
C.~J. Geyer, ``{Practical Markov chain Monte Carlo},'' \emph{Statistical
  science}, pp. 473--483, 1992.

\end{thebibliography}

\end{document}